\documentclass[aps,pra,onecolumn,prabib,nofootinbib]{revtex4}
\usepackage{graphicx}
\usepackage{amsmath}
\usepackage{amssymb,wasysym}
\usepackage{amsfonts}
\usepackage{bm,verbatim}

\newcommand{\bea}{\begin{eqnarray}}
\newcommand{\eea}{\end{eqnarray}}
\newcommand{\be}{\begin{equation}}
\newcommand{\ee}{\end{equation}}

\newcommand{\kk}{\mathbf{k}}

\newcommand{\KK}{\mathbf{K}}

\begin{document}

\title{Integral equations for the four-body problem}
\author{Christophe Mora${}^1$, Yvan Castin${}^2$,  Ludovic Pricoupenko${}^3$}
\affiliation{Laboratoires ${}^1$Pierre Aigrain and ${}^2$Kastler Brossel,
\'{E}cole Normale Sup\'{e}rieure and CNRS, UPMC${}^1$  and Paris 7 Diderot${}^2$, 24 rue Lhomond, 75231 Paris, France \\
${}^3$Laboratoire de Physique Th\'eorique de la Mati\`ere Condens\'ee,
UPMC and CNRS, 4 place Jussieu, 75005 Paris, France}
\date{\today}

\begin{abstract}
We consider the four-boson and 3+1 fermionic problems with a model Hamiltonian which encapsulates the mechanism of the Feshbach resonance involving the coherent coupling of two atoms in the open channel and a molecule in the closed channel. The model includes also the pair-wise direct interaction between atoms in the open channel and in the bosonic case, the direct molecule-molecule interaction in the closed channel. Interactions are modeled by separable potentials which makes it possible to reduce the four-body problem to the study of a single integral equation. We take advantage of the rotational symmetry and parity invariance of the Hamiltonian to reduce the general eigenvalue equation in each angular momentum
sector to an integral equation for functions of three real variables only. A first application of this formalism in the zero-range limit 
is given elsewhere [Y. Castin, C. Mora, L. Pricoupenko,  Phys. Rev. Lett. {\bf 105}, 223201 (2010)].
\vskip 0.5\baselineskip

\vskip 0.5\baselineskip
\end{abstract}

\maketitle

\section{Introduction}
\label{sec:intro}

The few-body problem is the object of a renewed interest \cite{revue_Efimov}, considering
the possibility to study experimentally this problem in a resonant regime with cold atoms 
close to a Feshbach resonance \cite{manips_Efimov}.
In this resonant regime, the $s$-wave scattering length $a$ describing the interaction between
the particles is  much larger (in absolute value) than the range $b$ of the interaction potential, 
which may be estimated by the van der Waals length of a few nanometers \cite{revue_feshbach2}.

The three-body resonant problem, initiated in particular in Refs.~\cite{Ter,Danilov,Efimov}, has been the subject of numerous
studies \cite{revue_Efimov}. Two distinct situations have been identified, depending on the quantum statistics 
and the mass ratios of the particles. In the first case, 
there exists a zero-range limit, where both
the true range $b$ and the effective range $r_e$ of the interaction tend to zero. The interactions may then
be replaced by contact conditions on the wavefunction, which define the so-called Bethe-Peierls \cite{Wigner,Bethe}
or zero-range model. The interaction is then described by $a$ as the single parameter.
The paradigm of this first case is three spin-1/2 fermions with equal
masses for the two spin states. In the second case, {\sl the Efimovian} case,
the zero-range limit does not exist and is replaced by a limit cycle.
The minimal model is then a modified zero-range model supplemented with three-body contact conditions involving the three-body
parameter $R_t$ that depends on the microscopic details of the interaction.
Then the interaction is described by only two parameters: $a$ and $R_t$. The archetype of the second
case is three indistinguishable bosons. The second case leads to the Efimov effect, that is for $|a|=\infty$ there
is an infinite number of weakly bound trimers with an accumulation point in the spectrum at zero energy.

In reality, the physics of the Feshbach resonance is not exhausted by the zero-range
model \cite{revue_feshbach2}. In particular, a Feshbach resonance involves two channels in the interaction
between two low-energy atoms, 
one channel being open and the other one being closed due to energy conservation.
In the specific case of ultra-narrow Feshbach resonances, where the effective range $r_e$ is negative and much larger
in absolute value than the true range $b$, $r_e$ becomes an important parameter.
It may be taken into account by modified Bethe-Peierls contact conditions \cite{Efimov_re,Petrov_re,Pricoupenko_re}, the
so-called effective range approach, or more
straightforwardly by use of a two-channel model \cite{Hussein,Holland,KoehlerBurnett,Koehler,Lee,Gurarie,ondep,Mora,Kokkelmans}.
In the general case, Feshbach resonances are  richer, as \emph{e.g.}\ a form or potential resonance may take
place in the open channel, in which case the background scattering length $a_{\rm bg}$ is
large and the modified Bethe-Peierls conditions \cite{Efimov_re,Petrov_re,Pricoupenko_re}
may not reproduce correctly the momentum dependence of the two-body scattering amplitude at low energy 
\footnote{E.g. at infinite scattering length $1/a=0$ and in the zero true range limit $b\to 0$, 
the modified Bethe-Peierls model accurately reproduces the two-channel model scattering amplitude $f_k$ of Eq.(\ref{eq:exprf})
over the physically relevant momentum range $k\lesssim k_{\rm typ}$ iff $|ik-\frac{1}{2} k^2 r_e+1/f_k|\ll |ik-\frac{1}{2} k^2 r_e|$\, 
$\forall k \in[0,k_{\rm typ}]$,
with $r_e=-2R_*$.  This is satisfied iff $\sup_{k\in [0,k_{\rm typ}]} k R_* |1-1/(1+k^2 a_{\rm bg} R_*)|/|i+k R_*| \ll 1$.
If $a_{\rm bg} = O(b)\to 0$, the supremum is zero.  If $a_{\rm bg}$ is so large and negative that  $k_{\rm typ} (-a_{\rm bg} R_*)^{1/2} >1$,
the supremum is infinite and the condition is violated.}
\cite{Kokkelmans,Marcelis,Stoof,Mattia}.
Using a two-channel model, with the description of the direct atomic interaction in the
open channel by a separable potential, is then the simplest way to have a reasonably  accurate
description of the atomic physics behind the Feshbach resonance, while keeping the simplicity
of the integral equations resulting from zero-range models.

The four-body problem is now under active investigation. The situation is more open,
since no analytical solution is available (contrary to the three-body case). 
In the absence of Efimov effect, the plain Bethe-Peierls model may be used for broad Feshbach resonances
\cite{Minlos,Petrov_dim_dim}. In principle exact numerical calculation of the resulting
integral equation may be done, and was indeed performed to determine a dimer-dimer scattering length
in Ref.~\cite{Petrov_dim_dim}. Also efficient numerical methods have been used
to solve the case of four or more harmonically trapped two-component fermions \cite{vonStecher,Bertsch,Blume}
In the Efimovian case, for four bosons,
finite range real space models, specified by an interaction potential $V(r_{ij})$,
have been recently used in the context of cold atoms but have been solved only in an approximate way \cite{Greene}.
As a consequence, the tetramer states predicted in \cite{Greene} that have (negative) energies larger
than the trimer ground state energy are actually resonances rather than genuine stationary states.
In the nuclear physics context, however, exact numerical solutions of the four-body problem with real space
interaction potentials $V(r_{ij})$ can be obtained thanks to the reformulation in terms of
momentum-space Faddeev-Yakubovsky integral equations \cite{FY1,FY2}.
A numerical solution of a low-energy effective theory based on this integral formulation  
was used in Ref.~\cite{Hammer_4corps} to predict universal properties of the ground four-boson state of that theory.
The Faddeev-Yakubovsky integral formulation may also be used to solve the four-body problem with
a separable potential interaction and to calculate the energy width of the resonances
corresponding to the tetramers of \cite{Greene}, as recently done in \cite{Deltuva}.
A limitation of the present Faddeev-Yakubovsky formulation seems to be, however, that it does not
encapsulate the atomic physics of the Feshbach resonance.

To answer as quantitatively as possible current questions on the four-body problem,
we use in this paper a four-parameter two-channel model to derive without approximation
a simple integral equation in momentum space for the four-body problem, involving an unknown function of two atomic momenta
only.  No solution of the equation is presented here, but we presented elsewhere a first application of this formalism in the limiting
case of a zero-range interaction \cite{Efimov4} and we expect that other applications will come.
In this paper, we take as specific examples first the case of four spinless bosons, in section \ref{sec:fbp},
and then the so-called $3+1$ fermionic problem of three same spin state fermions of mass $m$ interacting
with a distinguishable particle of mass $M$, in section \ref{sec:tpofp}.
The reduction of the integral equation thanks to the rotational symmetry of the Hamiltonian
is performed in section \ref{sec:tari}, leading to unknown functions of three real variables only. 
We conclude in section \ref{sec:conclusion}.

\section{The four-boson problem}
\label{sec:fbp}

\subsection{Model Hamiltonian and two-body scattering amplitude}
\label{subsec:model}

Our model is a two-channel model of a magnetic Feshbach resonance \cite{Hussein,Holland}.
This means that the atoms may actually exist in two different forms in the model,
either as atoms in the open channel, treated as elementary bosons, 
or as molecules in the closed channel, the so-called {\it closed channel} molecules.
Since these molecules have a size of the order of the van der Waals length $b$,
usually smaller than the mean distance between them, they are also treated 
as elementary bosons.
There exists a short-range coupling between the two channels, due in particular
to the existence of hyperfine atomic degrees of freedom, so that two atoms may 
coherently be converted to a closed-channel molecule, and {\sl vice versa}.
This coherent interconversion leads to an effective interaction between
the atoms, that may be made resonant by tuning with a magnetic field
the energy $E_{\rm mol}$ of a closed-channel molecule counted
with respect to the dissociation limit of the open channel potential.
Even in the absence of this interconversion, atoms in the open channel
can interact {\sl via} the van der Waals potential, and this so-called
background interaction is sometimes also resonant and cannot be neglected.

Our model Hamiltonian thus involves two coupled bosonic fields, one for
the open channel atoms and one for the closed-channel molecules. 
For simplicity, we assume that there is no trapping potential and we perform
calculations directly in momentum space without any quantization volume
(the real space version of the Hamiltonian was written in Ref.~\cite{Tarruell}).
The creation $a_\kk^\dagger$ and annihilation $a_\kk$ operators of one atom of
wave vector $\kk$,
and the creation $b_\kk^\dagger$ and annihilation $b_\kk$ operators of one closed
channel molecule of center-of-mass wave vector $\kk$ then obey the canonical commutation relations
\be
[a_\kk, a_{\kk'}^{\dagger}] = [b_\kk, b_{\kk'}^{\dagger}] = (2\pi)^3 \delta(\kk-\kk').
\ee
The other commutators are zero, in particular $a_\kk$ and $a^\dagger_\kk$ commute with $b_\kk$ and $b^\dagger_\kk$.
The Hamiltonian may be split in four terms,
\be
H = H_{\rm at} + H_{\rm mol} + H_{\rm open} + H_{\rm at-mol}.
\label{eq:hamilb}
\ee
The first two terms contain the kinetic energy (and the internal energy) of
non-interacting atoms and molecules,
\bea
H_{\rm at} &=& \int \frac{d^3k}{(2\pi)^3}  E_\kk a_\kk^\dagger a_\kk \\
H_{\rm mol} &=& \int \frac{d^3k}{(2\pi)^3}  (E_{\rm mol}+\frac{1}{2}E_\kk)  b_\kk^\dagger b_\kk.
\eea
We have introduced the convenient notation for the kinetic energy of an atom of wave vector $\kk$,
\be
E_\kk = \frac{\hbar^2 k^2}{2m},
\ee
and we shall use in what follows the parity invariance $E_{\kk}=E_{-\kk}$.

The third term in Eq.~(\ref{eq:hamilb}) describes the direct interaction between atoms
in the open channel. In principle, one should describe this interaction using true binary potentials
$V(r_{ij})$ with a $1/r_{ij}^6$ Van der Waals tail \cite{Moerdijk,Gao,Flambaum}. This however precludes
the derivation of a simple integral equation for the four-body problem. We thus rather take the convenient form 
of a separable potential, see e.g. \cite{Nozieres}:
\be
H_{\rm open} = \frac{g_0}{2}
\int \frac{d^3k_1}{(2\pi)^3} \frac{d^3k_2}{(2\pi)^3} \frac{d^3k_3}{(2\pi)^3} \frac{d^3k_4}{(2\pi)^3}
\\
\chi(\frac{\kk_3-\kk_4}{2}) \chi(\frac{\kk_2-\kk_1}{2})
(2\pi)^3 \delta(\kk_1+\kk_2-\kk_3-\kk_4) a_{\kk_3}^\dagger a_{\kk_4}^\dagger a_{\kk_2} a_{\kk_1}.
\ee
It involves a bare coupling constant $g_0$ and a cut-off function $\chi$ making the theory well defined.
In practice, we take as in Ref.~\cite{KoehlerBurnett,ondep} a real Gaussian cut-off function, which is both
a smooth and rapidly decreasing function of $k$,
\be
\chi(\kk) = e^{-k^2 b^2/2},
\ee
where $b$ is of the order of the van der Waals length. The Gaussian choice is of course arbitrary,
and may be replaced by any rapidly decreasing function of $k$. An important point is that the separable
potential is able to capture the essential features of the Feshbach resonance, such as the
dependance of the scattering length with the magnetic field (\ref{eq:aB}), 
or the structure (\ref{eq:regen}) of the effective range.

The last term in Eq.~(\ref{eq:hamilb}) describes the coherent interconversion of pairs
of atoms into closed-channel molecules, responsible for the Feshbach resonance:
\be
H_{\rm at-mol} = \Lambda \int  \frac{d^3k_1}{(2\pi)^3} \frac{d^3k_2}{(2\pi)^3}
\left[\chi(\frac{\kk_2-\kk_1}{2})b_{\kk_1+\kk_2}^\dagger a_{\kk_2} a_{\kk_1}
+ \chi(\frac{\kk_2-\kk_1}{2}) a_{\kk_1}^\dagger a_{\kk_2}^\dagger b_{\kk_1+\kk_2}\right]
\ee
It involves a coupling amplitude $\Lambda$ between the two channels, taken to be real,
and (for simplicity) the same cut-off
function $\chi$ of the relative momentum of two bosons, implementing in momentum space
the fact that, in real space, the conversion process is non-local over a radius
$\approx b$ only, that is two atoms are converted to a closed-channel molecule
with an appreciable probability amplitude only if their interatomic distance is of the order of $b$ or less.
This cut-off function is actually required to make $H_{\rm at-mol}$ mathematically sound.
Note that each of the four terms of the Hamiltonian conserves the total momentum
and preserves Galilean invariance, as it should be.

It appears that our model Hamiltonian
depends on four bare parameters $b$, $g_0$, $E_{\rm mol}$ and $\Lambda$. Apart from the true range
$b$, the other three parameters are more physically expressed in terms of the scattering length
$a$, the background scattering length $a_{\rm bg}$ and the always positive Feshbach length
$R_*$ \cite{Petrov_re}.  The link is performed thanks to the solution of the two-atom scattering problem,
that was detailed in \cite{Tarruell} for two opposite spin fermions, and that is adapted to the present 
bosonic case by replacing $\Lambda^2$ in the result of \cite{Tarruell} by $2\Lambda^2$.
The resulting expression of the two-body $T$ matrix is then 
\be
\label{eq:matriceT}
\langle \kk_f | T(E+i0^+) |\kk_i\rangle =
-\frac{4\pi\hbar^2}{m} \chi(\kk_f)\chi(\kk_i) \, f(E+i0^+)
\ee
where $\kk_i$ and $\kk_f$ are the wave vectors of the incoming wave and outgoing wave in the center of
mass frame, $E$ is the energy in that frame and we have defined the function of energy $f(E+i0^+)$ such that:
\be
\frac{-m}{4\pi\hbar^2 f(E+i0^+)} =
\left[g_0 +\frac{2\Lambda^2}{E-E_{\rm mol}}\right]^{-1} -\int\frac{d^3k}{(2\pi)^3} \frac{\chi^2(\kk)}{E+i 0^+-2E_\kk}.
\ee
The calculation in \cite{Tarruell} was done on shell, that is for $k_i^2=k_f^2=mE/\hbar^2$, and $E\geq 0$.
In this case, the function $f(E)$ is simply related to the $s$-wave scattering amplitude $f_k$ at that energy,
with $k=(mE)^{1/2}/\hbar$, by the relation
\be
\label{eq:fk_vs_fE}
f_k = \chi(\kk_f)\chi(\kk_i)f(E+i0^+).
\ee
By an analytic continuation argument,
Eq.~(\ref{eq:matriceT}) actually also holds
off shell, for all values of $\kk_i$, $\kk_f$ and $E$.
Assuming that $E_{\rm mol}$ is an affine function of the magnetic field $B$,
a good approximation if $B$ remains close to the Feshbach resonance location $B_0$,
one extracts from Eq.~(\ref{eq:matriceT}) the standard formula for the magnetic field
dependence of the scattering length $a\equiv \lim_{k\to 0} -f_k$,
\be
a = a_{\rm bg} \left[ 1 -\frac{\Delta B}{B-B_0}\right],
\label{eq:aB}
\ee
where $a_{\rm bg}$ is the background scattering length and $\Delta B$ is the width
of the Feshbach resonance. The Feshbach length is connected to this width by
\be
R_* = \frac{\hbar^2}{m a_{\rm bg} \delta\mu \Delta B},
\ee 
where $\delta\mu = dE_{\rm mol}/dB$. The physical parameters of the Feshbach resonance
are finally found to be related to the bare ones by
\bea
\label{eq:phys1}
\frac{1}{a_{\rm bg}} &=& \frac{4\pi\hbar^2/m}{g_0} + \frac{1}{b\sqrt{\pi}} \\
\label{eq:phys2}
\frac{1}{a} &=& \frac{4\pi\hbar^2/m}{g_0-2\Lambda^2/E_{\rm mol}} + \frac{1}{b\sqrt{\pi}} \\
\label{eq:phys3}
R_* &=& \frac{g_0^2}{8\pi\Lambda^2 a_{\rm bg}^2},
\eea
which allows a direct connection with the experimental parameters.
The expression of the function $f(E)$ of our model for $E\geq 0$, and of its analytic continuation
to $E<0$, is also quite simple in terms of the physical
parameters \cite{Tarruell},
\be
\label{eq:exprf}
-\frac{e^{k^2 b^2}}{f(E+i0^+)} = -\frac{1}{f_k} =
\frac{e^{k^2b^2}}{a}\left[1-\left(1-\frac{a}{a_{\rm bg}}\right)\frac{k^2}{k^2-Q^2}\right]-i k\, \mathrm{erf}\,(-i k b) + i k
\ee
where $\mathrm{erf}$ is the error function, 
we have set $k=(mE)^{1/2}/\hbar$ for $E\geq 0$ and $k=i(-mE)^{1/2}/\hbar$ for $E<0$, and we have introduced
$Q^2$ such that $-1/Q^2=a_{\rm bg} R_* (1-a_{\rm bg}/a).$
This allows in particular a direct calculation of the effective range $r_e$ of the model \cite{Tarruell}, defined as
$-1/f_k=\frac{1}{a}+ik -\frac{1}{2} k^2 r_e + o(k^2)$ for $k\to 0$,
\be
r_e = -2 R_* \left(1-\frac{a_{\rm bg}}{a}\right)^2 +\frac{4b}{\sqrt{\pi}}-\frac{2b^2}{a}.
\label{eq:regen}
\ee
Another quantity of interest for what follows is the internal energy $E_{\rm mol}$ of a decoupled closed-channel molecule.
It may be expressed in terms of $b$ and of the three physical parameters $a,a_{\rm bg}$ and $R_*$
using Eqs.~(\ref{eq:phys1},\ref{eq:phys2},\ref{eq:phys3}). For simplicity, we give this expression
only at resonance $1/a=0:$
\be
E_{\rm mol}^{\rm res} =  -\frac{\hbar^2}{mR_*(a_{\rm bg} - b \sqrt{\pi})} .
\label{eq:Emol_res}
\ee
Since $R_*$ is positive, $E_{\rm mol}^{\rm res}$ is negative if and only if $a_{\rm bg} > b \sqrt{\pi}$.

\subsection{Four-body ansatz and coupled equations}

We consider for simplicity a four-body eigenvector of our model Hamiltonian in free space
with an eigenenergy $E$ obeying the two conditions
\be
\label{eq:cond_sur_E}
E < 0 \ \ \ \ \mbox{and} \ \ \ \   E - 2E_{\rm mol}   < 0.
\ee
The case of $E>0$ is a four-atom scattering problem that
may be treated along the lines of \emph{e.g.}\ \cite{ondep}, including in particular an 
incoming free four-atom state. For the same reason of simplicity, to avoid the treatment
of the scattering problem of two closed-channel molecules, we also impose
$E-2 E_{\rm mol} <0$
\footnote{As a reminder, the solution of a free space problem involves two aspects, Schr\"odinger's equation itself
and boundary conditions that one must impose on the solution to recover the right physics. Implementing these boundary conditions
is performed at two levels: (i) in the derivation of integral equations from Schr\"odinger's equation,
when one divides by an energy difference $E-E_{\rm free}$, where $E_{\rm free}$ is usually the unperturbed
energy of some configuration, e.g. plane waves, one has to be careful if $E-E_{\rm free}$ can be zero, in which case the
resulting vanishing denominators have to be regularized by the inclusion of a vanishing, usually positive imaginary part, which amounts
to replacing $E$ with $E+i0^+$; this division by a vanishing quantity also means that an additive free wave solution has
to be included, that corresponds physically to an incoming wave. (ii) Even if all energy denominators
$E-E_{\rm free}$ are strictly negative, as it is the case in this paper, one in general still has to choose an ansatz for the unknown
functions [here the functions $A, B, C$] that may be the sum of an incoming bit and of a scattered bit. E.g. the scattering 
of two dimers of opposite incoming wavevectors $\pm\kk_0$ may take place at total negative energy, 
so that the integral equations derived here strictly hold [level (i) may be skipped],
but still ansatz have to be given, $C(\mathbf{k})=\frac{1}{2}(2\pi)^3 [\delta(\mathbf{k}-\mathbf{k}_0)+\delta(\mathbf{k}+\mathbf{k}_0)]
p_{\rm closed}+C_{\rm scatt}(\mathbf{k}),$ where $p_{\rm closed}$ is the probability for a dimer
to be in the closed channel and $C_{\rm scatt}(\mathbf{k})$ is the scattered wave contribution, etc.}
In the two-channel model, the four-body state vector is in general a coherent superposition
of a component with four atoms in the open channel, a component with two atoms in the open channel
and one closed-channel molecule, and a component with two closed-channel molecules,
\be
|\Psi\rangle = |\psi_{4\,\rm at}\rangle + |\psi_{2\,\rm at+1\,mol}\rangle +  |\psi_{\rm 2\,mol}\rangle.
\label{eq:ansatzb}
\ee
Restricting without loss of generality to a zero total momentum state,  we obtain the ansatz
\bea
|\psi_{4\,\rm at}\rangle &=&
\int \frac{d^3k_1}{(2\pi)^3} \frac{d^3k_2}{(2\pi)^3} \frac{d^3k_3}{(2\pi)^3} \frac{d^3k_4}{(2\pi)^3}
(2\pi)^3 \delta(\kk_1+\kk_2+\kk_3+\kk_4)
A(\kk_1,\kk_2,\kk_3,\kk_4) a_{\kk_1}^\dagger a_{\kk_2}^\dagger a_{\kk_3}^\dagger a_{\kk_4}^\dagger |0\rangle, \\
|\psi_{2\,\rm at+1\,mol}\rangle &=&  \int \frac{d^3k_1}{(2\pi)^3} \frac{d^3k_2}{(2\pi)^3}
B(\kk_1,\kk_2) b_{-(\kk_1+\kk_2)}^\dagger a_{\kk_1}^\dagger a_{\kk_2}^\dagger |0\rangle, \\
|\psi_{\rm 2\,mol}\rangle &=& \int \frac{d^3k}{(2\pi)^3} C(\kk) b_\kk^\dagger b_{-\kk}^\dagger |0\rangle.
\eea
Thanks to the bosonic symmetry, ensuring that all the creation operators commute,
we can impose that the amplitude $A$ is a symmetric function of its four arguments,
the amplitude $B$ is a symmetric function of its two arguments, and $C(\kk)$ is an even function of
$\kk$.

It remains to calculate $H |\Psi\rangle$ and to project Schr\"odinger's equation $0= (H-E) |\Psi\rangle$ 
on the three relevant subspaces. Projection in the sector with four open channel atoms
yields
\be
\label{eq:0=F4}
0=\int \frac{d^3k_1}{(2\pi)^3} \frac{d^3k_2}{(2\pi)^3} \frac{d^3k_3}{(2\pi)^3} \frac{d^3k_4}{(2\pi)^3}
(2\pi)^3 \delta(\kk_1+\kk_2+\kk_3+\kk_4)
F_4(\kk_1,\kk_2,\kk_3,\kk_4)  a_{\kk_1}^\dagger a_{\kk_2}^\dagger a_{\kk_3}^\dagger a_{\kk_4}^\dagger |0\rangle
\ee
with the function
\be
\label{eq:F4}
F_4(\kk_1,\kk_2,\kk_3,\kk_4) = \left[-E+\sum_{n=1}^{4} E_{\kk_n}\right] A(\kk_1,\kk_2,\kk_3,\kk_4)
+ \chi(\frac{\kk_1-\kk_2}{2}) [6 g_0 \tilde{B}(\kk_3,\kk_4) +\Lambda B(\kk_3,\kk_4)].
\ee
The first term in the right-hand side clearly contains the kinetic energies of the four atoms.
In the second term, the occurrence of the $B$ amplitude of the ansatz is due to the conversion
of a closed-channel molecule into two open channel atoms, as is revealed by the presence
of the factor $\Lambda$. Finally, the term proportional to $g_0$ is due to the direct pair-wise
interaction between the four atoms in the open channel. Since this interaction is modeled by 
a separable potential, it involves a function of two momenta only, obtained by contracting
the four-boson amplitude $A$ with the cut-off function $\chi$,
\be
\label{eq:Bt}
\tilde{B}(\kk_1,\kk_2) \equiv \int \frac{d^3k_3}{(2\pi)^3} \frac{d^3k_4}{(2\pi)^3}
(2\pi)^3 \delta(\kk_1+\kk_2+\kk_3+\kk_4)
\chi(\frac{\kk_3-\kk_4}{2}) A(\kk_3,\kk_4,\kk_1,\kk_2).
\ee
We have introduced the notation $\tilde{B}$ to underline the formal similarity of this term
in Eq.~(\ref{eq:F4}) with the two-channel contribution $\Lambda B$.
Eqs.~(\ref{eq:0=F4},\ref{eq:F4}) allow to express the four-boson amplitude $A$ in terms of two-body amplitudes
$B$ and $\tilde{B}$. Because of bosonic symmetry, Eq.~(\ref{eq:0=F4}) does not imply that
$F_4$ is identically zero, it only implies that the totally symmetric component of $F_4$,
symmetrized over the $4!$ permutations $\sigma$ of the four arguments $\kk_{1,2,3,4}$, is zero.
This leads to
\be
\label{eq:A}
A(\kk_1,\kk_2,\kk_3,\kk_4) = \frac{1}{4!} \sum_{\sigma\in S_4}
\chi(\frac{\kk_{\sigma(1)}-\kk_{\sigma(2)}}{2})
\frac{6 g_0 \tilde{B}(\kk_{\sigma(3)},\kk_{\sigma(4)}) +\Lambda B(\kk_{\sigma(3)},\kk_{\sigma(4)})}
{E-\sum_{n=1}^{4} E_{\kk_n}}.
\ee
The unknown amplitude $A$ can thus be eliminated in the remaining part of the calculation.
Insertion of Eq.~(\ref{eq:A}) in the definition (\ref{eq:Bt}) gives
\be
\tilde{B}(\kk_1,\kk_2) =
\int \frac{d^3k_3}{(2\pi)^3}\frac{d^3k_4}{(2\pi)^3}
\frac{(2\pi)^3 \delta(\sum_{n=1}^4\kk_n)}{E-\sum_{n=1}^{4} E_{\kk_n}}
\frac{1}{4!} \sum_{\sigma\in S_4}
\chi(\frac{\kk_3-\kk_4}{2}) \chi(\frac{\kk_{\sigma(1)}-\kk_{\sigma(2)}}{2})
[6 g_0 \tilde{B}+\Lambda B](\kk_{\sigma(3)},\kk_{\sigma(4)}).
\ee
Collecting permutations giving identical contributions, up to a relabeling of the integration variables,
gives the more explicit form 
\footnote{
One can distinguish three types of permutations. Type one includes four permutations:
It corresponds to $\sigma\{1,2\}=\{3,4\}$ (which implies $\sigma\{3,4\}=\{1,2\}$). These four permutations
give equal contributions. Type two
also includes four permutations: It corresponds to $\sigma\{1,2\}=\{1,2\}$ (which implies $\sigma\{3,4\}=\{3,4\}$).
These four permutations give equal contributions.
The third type includes sixteen permutations: It corresponds to the cases where
there exists a unique $i\in\{1,2\}$ such that $\sigma(i)\in\{3,4\}$ (which implies that there exists
a unique $j\in\{3,4\}$ such that $\sigma(j)\in\{1,2\}$). One can reduce to the case
$\sigma(i)=3$, exchanging the integration variables $\kk_3$ and $\kk_4$ if necessary.
Let $i'$ be the element of $\{1,2\}$ different from $i$.
Then two cases remain to be distinguished, the case $\sigma(i')=1$ and the case $\sigma(i')=2$,
which give different contributions.
}
\bea
\tilde{B}(\kk_1,\kk_2) &=& \int \frac{d^3k_3}{(2\pi)^3} \frac{d^3k_4}{(2\pi)^3}
\frac{(2\pi)^3 \delta(\sum_{n=1}^{4} \kk_n)}{E-\sum_{n=1}^{4} E_{\kk_n}}
\, \frac{1}{6} \chi(\frac{\kk_3-\kk_4}{2})
\left\{\chi(\frac{\kk_3-\kk_4}{2}) [6 g_0 \tilde{B}+\Lambda B](\kk_1,\kk_2)
\right.
\nonumber \\
&& \left.
+ \chi(\frac{\kk_1-\kk_2}{2})
[6 g_0 \tilde{B}+\Lambda B](\kk_3,\kk_4)
+2 \left[
\chi(\frac{\kk_3-\kk_1}{2}) [6 g_0 \tilde{B}+\Lambda B](\kk_2,\kk_4)
+1\leftrightarrow 2
\right]
\right\}.
\label{eq:imp1}
\eea
This equation is the first important equation of this subsection.

We now project Schr\"odinger's equation $0= (H-E) |\Psi\rangle$ in the sector with two open channel atoms
and one closed-channel molecule. We obtain
\be
\label{eq:0=F2}
0= \int \frac{d^3k_1}{(2\pi)^3} \frac{d^3k_2}{(2\pi)^3} F_2(\kk_1,\kk_2)
b_{-(\kk_1+\kk_2)}^\dagger a_{\kk_1}^\dagger a_{\kk_2}^\dagger |0\rangle
\ee
with
\bea
F_2(\kk_1,\kk_2)&=&[E_{\kk_1}+E_{\kk_2}+\frac{1}{2} E_{\kk_1+\kk_2} + E_{\rm mol}-E] B(\kk_1,\kk_2)
+12\Lambda \tilde{B}(\kk_1,\kk_2) \nonumber \\
&+& \chi(\frac{\kk_1-\kk_2}{2})[2\Lambda C(\kk_1+\kk_2)+ g_0 \beta(\kk_1+\kk_2)].
\label{eq:F2}
\eea
In the first term in the right-hand side of $F_2$, we recognize the kinetic energies of two atoms,
plus the kinetic and internal energies of a closed-channel molecule of wavevector $-(\kk_1+\kk_2)$.
The contributions $\Lambda \tilde{B}$ and $\Lambda C$ are due to the interchannel coupling, respectively
converting a pair of atoms into a closed-channel molecule and {\sl vice versa}. The fact that
the same contraction $\tilde{B}$ defined in Eq.~(\ref{eq:Bt}) appears as in the four-atom sector is due
to the choice of the same cut-off function in the separable potential and the open-to-closed-channel
coupling. Finally, the term proportional to $g_0$ in Eq.~(\ref{eq:F2}) results from the direct open channel
interaction between the two open channel atoms present in $|\psi_{2\,\rm at+1\,mol}\rangle$. This involves
the contraction of $B$ with the cut-off function $\chi$, that we call $\beta$,
\be
\label{eq:defbeta}
\beta(\KK) = \int \frac{d^3k_1}{(2\pi)^3} \frac{d^3k_2}{(2\pi)^3} (2\pi)^3 \delta(\kk_1+\kk_2-\KK)
\chi(\frac{\kk_1-\kk_2}{2}) B(\kk_1,\kk_2).
\ee
Since $F_2$ is a symmetric function of $\kk_1$ and $\kk_2$, Eq.~(\ref{eq:0=F2}) directly imposes
$F_2(\kk_1,\kk_2)=0.$

Finally, projecting Schr\"odinger's equation $0=(H-E)|\Psi\rangle$ in the sector with zero open channel atom
and two closed-channel molecules gives
\be
\label{eq:0=F0}
0 = \int \frac{d^3k}{(2\pi)^3} F_0(\kk) b_\kk^\dagger b_{-\kk}^\dagger |0\rangle
\ee
with
\be
F_0(\kk)= (2E_{\rm mol}+E_\kk-E) C(\kk) + \Lambda [\beta(\kk)+\beta(-\kk)].
\ee
The first term in the right-hand side contains the kinetic and internal energies of the two
closed-channel molecules, and the last term originates from the conversion of two atoms
into a closed-channel molecule. The direct open channel interaction does not enter here.
The fact that the same contracted function $\beta$ appears as in Eq.~(\ref{eq:F2}) is due to the
fact that the same cut-off function $\chi$ is used in $H_{\rm open}$ and  $H_{\rm at-mol}$.
Since $F_0(\kk)$ is an even function of $\kk$, Eq.~(\ref{eq:0=F0}) directly imposes $F_0(\kk)=0$,
which allows to eliminate $C$,
\be
C(\kk) = \Lambda \frac{\beta(\kk)+\beta(-\kk)}{E-E_\kk-2E_{\rm mol}}.
\ee
We used the condition Eq.~(\ref{eq:cond_sur_E}) to ensure that the energy denominator in that
expression cannot vanish.
Reporting this into the equation $F_2=0$ gives the second important equation of this subsection,
\bea
-12\Lambda \tilde{B}(\kk_1,\kk_2) &=&
[E_{\kk_1}+E_{\kk_2}+\frac{1}{2} E_{\kk_1+\kk_2} + E_{\rm mol}-E] B(\kk_1,\kk_2)
\nonumber \\
&+&\chi(\frac{\kk_1-\kk_2}{2})\left[ g_0 \beta(\kk_1+\kk_2)
+2\Lambda^2 \frac{\beta(\kk_1+\kk_2)+\beta(-\kk_1-\kk_2)}{E-E_{\kk_1+\kk_2}-2E_{\rm mol}}
\right].
\label{eq:imp2}
\eea

\subsection{A single integral equation}

We now show that, remarkably, the rather involved set of equations (\ref{eq:imp1}) and (\ref{eq:imp2})
can be reduced to a single explicit integral equation on the unknown function $D(\kk_1,\kk_2)$
defined as:
\be
\label{eq:defD}
\Lambda D(\kk_1,\kk_2) = 6 g_0\tilde{B}(\kk_1,\kk_2) + \Lambda B(\kk_1,\kk_2).
\ee
We note that this function naturally appears in the right-hand side of Eq.~(\ref{eq:imp1}). 
The left-hand side of Eq.~(\ref{eq:imp1}) involves $\tilde{B}$, which is a linear combination
of the functions $D$ and $B$. The only non trivial step is thus to express 
$B$ as a function of $D$ using Eq.~(\ref{eq:imp2}). We rewrite Eq.~(\ref{eq:imp2}) introducing
an effective $\beta$ function,
\be
\label{eq:defbeff}
\beta_{\rm eff}(\KK) \equiv \beta(\KK) + \frac{2\Lambda^2}{g_0} \frac{\beta(\KK)+\beta(-\KK)}{E-E_\KK-2 E_{\rm mol}},
\ee
and introducing what we call the relative energy for a reason that will become clear in the final result,
\be
E_{\rm rel}(\kk_1,\kk_2) \equiv E-E_{\kk_1}-E_{\kk_2}-\frac{1}{2} E_{\kk_1+\kk_2}.
\ee
Eliminating $\tilde{B}$ in terms of $D$ and $B$ thanks to Eq.~(\ref{eq:defD}), we transform Eq.~(\ref{eq:imp2}) into
\be
\label{eq:imp2bis}
\frac{D(\kk_1,\kk_2)}{u(\kk_1,\kk_2)} = B(\kk_1,\kk_2) -\frac{g_0^2}{2\Lambda^2}
\chi(\frac{\kk_1-\kk_2}{2}) \frac{\beta_{\rm eff}(\kk_1+\kk_2)}{u(\kk_1,\kk_2)}
\ee
with the notation
\be
u(\kk_1,\kk_2) \equiv 1- \frac{g_0}{2\Lambda^2}[E_{\rm mol}-E_{\rm rel}(\kk_1,\kk_2)].
\ee
Eq.~(\ref{eq:imp2bis}) does not immediately give $B$ in terms of $D$ because of the occurrence of $\beta_{\rm eff}$
in the right-hand side. Consequently, one has first to express $\beta_{\rm eff}$ in terms of $D$. We note that
the Hamiltonian $H$ of the system is invariant by parity. We can thus without loss of generality
assume that the eigenstate $|\Psi\rangle$ of $H$ that we consider is of well defined parity $\eta$,
with $\eta=1$ (even state) or $-1$ (odd state). Hence, the functions $B$, $\tilde{B}$ and thus $D$, $\beta$  have also
the parity $\eta$. For example,
\be
D(-\kk_1,-\kk_2) = \eta D(\kk_1,\kk_2) \ \ \ \mbox{and} \ \ \ \beta(-\KK)=\eta \beta(\KK).
\ee
This makes Eq.~(\ref{eq:defbeff}) local in $\KK$, expressing $\beta_{\rm eff}(\KK)$ as the product
of $\beta(\KK)$ and of a known $\eta$-dependent function of $\KK$.
Since $\beta(\KK)$ results from the contraction of $B(\kk_1,\kk_2)$  with the cut-off function
$\chi$, see Eq.~(\ref{eq:defbeta}), it remains to multiply Eq.~(\ref{eq:imp2bis})
by $\delta(\kk_1+\kk_2-\KK)\chi\left(\frac{\kk_1-\kk_2}{2}\right)$ and to integrate over $\kk_1$ and $\kk_2$.
In the left-hand side of Eq.~(\ref{eq:imp2bis}) the contraction of $D/u$ with $\chi$ immediately appears,
we call it
\be
\Delta(\KK) \equiv \int \frac{d^3k_1}{(2\pi)^3} \frac{d^3k_2}{(2\pi)^3} (2\pi)^3 \delta(\kk_1+\kk_2-\KK) 
\chi(\frac{\kk_1-\kk_2}{2})
\frac{D(\kk_1,\kk_2)}{u(\kk_1,\kk_2)}.
\ee
We finally obtain the desired expression $\beta_{\rm eff}$ in terms of $D$,
\be
\label{eq:desired_expression}
\beta_{\rm eff}(\KK) = v(\KK) \Delta(\KK)
\ee
where we have introduced the parity dependent notation
\be
v(\KK) = \left\{\left[1+\frac{(1+\eta)2\Lambda^2/g_0}{E-E_\KK-2 E_{\rm mol}}\right]^{-1} 
- \frac{g_0^2}{2\Lambda^2} \int\frac{d^3k}{(2\pi)^3} \frac{\chi^2(\kk)}{u(\frac{1}{2}\KK-\kk,\frac{1}{2}\KK+\kk)}\right\}^{-1}.
\ee
Eq.~(\ref{eq:imp2bis}) thus now gives $B$ explicitly as a functional of $D$. This, combined with Eq.~(\ref{eq:defD}),
immediately gives $\tilde{B}$ as a functional of $D$. The left-hand side of Eq.~(\ref{eq:imp1}) is then also
expressed in terms of $D$. Since the right-hand side of Eq.~(\ref{eq:imp1}) is directly a functional of
$D$, Eq.~(\ref{eq:imp1}) (multiplied for convenience by the factor $6/\Lambda$)
reduces to the desired closed integral equation for the unknown function $D$.
Collecting in the right-hand side of that integral equation the two terms involving $D(\kk_1,\kk_2)$, that is
one term originating from $\tilde{B}$ and the other term originating from the right-hand side
of Eq.~(\ref{eq:imp1}), we get as a prefactor of $D(\kk_1,\kk_2)$ the function
\be
\mbox{prefactor} = -\frac{1}{g_0} \left(1-\frac{1}{u(\kk_1,\kk_2)}\right)
+ \int \frac{d^3k_3}{(2\pi)^3} \frac{d^3k_4}{(2\pi)^3} \frac{(2\pi)^3 \delta(\sum_{n=1}^{4} \kk_n)}{E-\sum_{n=1}^{4} E_{\kk_n}} \\
\chi^2(\frac{\kk_3-\kk_4}{2}).
\ee
This may be transformed using the explicit expression of the function $u$, the identity
\be
\frac{(2\pi)^3 \delta(\sum_{n=1}^{4} \kk_n)}{E-\sum_{n=1}^{4} E_{\kk_n}} =
\frac{(2\pi)^3 \delta(\kk_1+\kk_2+\kk_3+\kk_4)}{E_{\rm rel}(\kk_1,\kk_2) - 2 E_{(\kk_4-\kk_3)/2}}
\ee
and performing the change of variable of unit Jacobian, $\kk_{3,4}=\frac{\KK}{2}\pm \kk$. Remarkably,
the prefactor can then be linked to the function $f$ appearing in the two-body $T$ matrix at the 
energy $E_{\rm rel}$, 
\be
\mbox{prefactor} = \frac{m}{4\pi\hbar^2 f[E_{\rm rel}(\kk_1,\kk_2)]}.
\ee
This justifies the name {\sl relative} energy for $E_{\rm rel}$. To conclude, we give the final
equation for $D$, which is the main result of this section:

\hspace{-6.5mm}\fbox{
\parbox{\textwidth}{
\bea
0 &=& \frac{m D(\kk_1,\kk_2)}{4\pi\hbar^2 f[E_{\rm rel}(\kk_1,\kk_2)]} +\frac{g_0}{2\Lambda^2} \chi(\frac{\kk_1-\kk_2}{2})
\frac{v(\kk_1+\kk_2)}{u(\kk_1,\kk_2)} \Delta(\kk_1+\kk_2) 
+\int \frac{d^3k_3}{(2\pi)^3} \frac{d^3k_4}{(2\pi)^3}
\Big\{ \frac{(2\pi)^3 \delta(\sum_{n=1}^{4}\kk_n)}{E-\sum_{n=1}^{4} E_{\kk_n}}
\nonumber\\
&&\chi(\frac{\kk_3-\kk_4}{2}) \left[\chi(\frac{\kk_1-\kk_2}{2}) D(\kk_3,\kk_4)
+2\chi(\frac{\kk_3-\kk_1}{2}) D(\kk_2,\kk_4) +
2\chi(\frac{\kk_3-\kk_2}{2}) D(\kk_1,\kk_4) \right]\Big\}.
\label{eq:forme_jolie}
\eea
}}
This has to be solved for a bosonic symmetry of $D$, $D(\kk_1,\kk_2)=D(\kk_2,\kk_1)$.
We note that the second term in the right-hand side of Eq.~(\ref{eq:forme_jolie}) has no equivalent
in zero-range theories and is to our knowledge new.

\subsection{Including closed-channel molecule interaction}

In the previous subsections, any direct interaction between closed-channel molecules
was neglected. Here we derive a modified closed integral equation for the function
$D(\mathbf k_1,\mathbf k_2)$, including the intermolecular $s$-wave interaction in the form of a separable potential
with the previously introduced cut-off function $\chi(\kk)$ and a bare coupling constant $g_0^{\rm mol}$.
This amounts to adding to the model Hamiltonian Eq.~(\ref{eq:hamilb}) the term
\begin{equation}
H_{\rm 2 mol} =\frac{1}{2} g_0^{\rm mol} 
\int \frac{d^3K}{(2\pi)^3} \frac{d^3k}{(2\pi)^3} 
\frac{d^3k'}{(2\pi)^3} \, 
\chi(\kk)\chi(\kk')
b^\dagger_{\frac{1}{2} \mathbf K+\mathbf k'} b^\dagger_{\frac{1}{2} \mathbf K-\mathbf k'} 
b_{\frac{1}{2}\mathbf K+\mathbf k} b_{\frac{1}{2} \mathbf K-\mathbf k}.
\label{eq:H2mol}
\end{equation}
This term $H_{\rm 2 mol}$ has a non-zero action only on the component
$|\psi_{\rm 2\,mol}\rangle$ of the four-body state vector $|\Psi\rangle$ in Eq.~(\ref{eq:ansatzb}).
Projecting Schr\"odinger's equation in that two closed-channel molecule sector gives Eq.~(\ref{eq:0=F0})
with the function $F_0(\KK)$ modified as
\begin{equation}
F_0(\mathbf K) = (2E_{\rm mol} + E_{\mathbf K} -E) C(\mathbf K) + \Lambda\left[\beta(\mathbf K) + \beta(-\mathbf K)\right]
+ g_0^{\rm mol} \int \frac{d^3K'}{(2\pi)^3} \chi(\KK) \chi(\KK')  C(\mathbf K').
\label{eq:2molsector}
\end{equation}
Since $F_0$ is an even function, Schr\"odinger's equation requires $F_0(\KK)=0$. This leads to an
integral equation of the Lippmann-Schwinger type:
\begin{equation}
C(\mathbf K) = C_0(\mathbf K)  + g_0^{\rm mol} G_{\rm mol}^0(\mathbf K) \chi(\mathbf K) 
\int \frac{d^3K'}{(2\pi)^3}  \chi(\mathbf K') C(\mathbf K'),
\label{eq:Lippman-2mol}
\end{equation}
where 
$C_0(\mathbf K) = \Lambda G^0_{\rm mol}(\mathbf K) \left[ \beta(\mathbf K) + \beta(-\mathbf K) \right],$
and ${G_{\rm mol}^0(\mathbf K)}=1/(E^{\rm rel}_{\rm mol}-E_{\mathbf k})$ 
is the free Green's function of the two closed-channel molecules
at the relative energy ${E^{\rm rel}_{\rm mol}}\equiv E-2E_{\rm mol}$.
We recall that here $E<2 E_{\rm mol}$, see Eq.~(\ref{eq:cond_sur_E}).
To solve the Lippmann-Schwinger equation (\ref{eq:Lippman-2mol}), one multiplies it by
$\chi(\KK)$ and one integrates over $\KK$. A closed equation is then obtained
for $\gamma\equiv \int \frac{d^3K}{(2\pi)^3} \chi(\KK) C(\KK).$ This leads
to the expression of ${C(\mathbf K)}$ as a functional of ${C_0(\mathbf K)}$,  
and thus of the contracted pair function $\beta(\mathbf K)$:
\begin{equation}
C(\mathbf K) = C_0(\mathbf K)  + g^{\rm mol} G_{\rm mol}^0(\mathbf K) \chi(\mathbf K) 
\int \frac{d^3K'}{(2\pi)^3} \chi(\mathbf K') C_0(\mathbf K'),
\end{equation}
where $g^{\rm mol}$ is the effective molecular coupling constant defined by
\begin{equation}
g^{\rm mol} \equiv \left(\frac{1}{g_0^{\rm mol}} 
- \int \frac{d^3K'}{(2\pi)^3} G^0_{\rm mol}(\mathbf K') \chi^2(\mathbf K') \right)^{-1} .
\end{equation}

Similarly to the previous subsection, 
the four-body problem at an energy satisfying Eq.~(\ref{eq:cond_sur_E})
can be reduced to an integral equation for the function $D(\kk_1,\kk_2)$ defined
in Eq.~(\ref{eq:defD}). Remarkably, one finds that the only change to perform to Eq.~(\ref{eq:forme_jolie})
is the substitution
\be
v(\mathbf K) \Delta(\mathbf K)  \to  v(\mathbf K) \Delta(\mathbf K) +\frac{g^{\rm mol} \chi(\mathbf K) v(\mathbf K) }
{1+\left[\frac{2(1+\eta)\Lambda^2}{g_0} G_{\rm mol}^0(\mathbf K)\right]^{-1}} 
\Bigg\{
\frac{-\frac{2\Lambda^2}{g_0^2} \int \frac{d^3K'}{(2\pi)^3}  \frac{\chi(\KK') G^0_{\rm mol}(\KK')\Delta(\KK')}
{R(\KK')I(\KK')}}
{1 +
\frac{2(1+ \eta) \Lambda^2 g^{\rm mol}}{g_0} \int \frac{d^3K'}{(2\pi)^3} \frac{[\chi(\KK') G^0_{\rm mol}(\KK')]^2}
{R(\KK')}}
\Bigg\}
\label{eq:subst}
\ee
where we have introduced the functions: 
\begin{equation}
R(\KK')=1+\frac{2(1+\eta)\Lambda^2}{g_0} G^0_{\rm mol}(\KK')-
\frac{2\Lambda^2}{g_0^2 I(\KK')} \ \ \mbox{and} \ \ I(\KK')=\int \frac{d^3k}{(2\pi)^3}
\frac{\chi^2(\mathbf k)}{{u(\frac{1}{2} \mathbf K' -\mathbf k,\frac{1}{2} \mathbf K' +\mathbf k)}}.
\end{equation}
As expected, the odd states ${(\eta=-1)}$ are not affected by the $s$-wave closed-channel molecule interaction.
For even states ${(\eta=1)}$, the term added by the substitution is in general non zero.
It may significantly affect the integral equation for $D$, that is even for the low-energy eigenstates
$|E|\ll \hbar^2/(mb^2)$, when the following resonance condition is met:
\begin{equation}
\left| 1+\frac{4\Lambda^2g^{\rm mol}}{g_0}\int \frac{d^3K'}{(2\pi)^3} \frac{[\chi(\mathbf K') G_{\rm mol}^0(\mathbf K')]^2}{R(\mathbf K')}\right| \ll 1 .
\label{eq:resonance_4b}
\end{equation}

To obtain Eq.~(\ref{eq:subst}), we had to recalculate the function $\beta_{\rm eff}\equiv \beta(\mathbf K) + 
(2\Lambda/g_0) C(\mathbf K)$, now including the closed-channel molecule interaction.
For this purpose, we used the expression of $\Delta$ in terms of $\beta$  and ${\beta_{\rm eff}}$:
\begin{equation}
\Delta(\mathbf K)=\beta(\mathbf K)- \beta_{\rm eff}(\mathbf K) \times \frac{g_0^2}{2\Lambda^2}\int\frac{d^3k}{(2\pi)^3} 
\frac{\chi^2(\mathbf k)}{u(\frac{1}{2} \mathbf K -\mathbf k,\frac{1}{2} \mathbf K +\mathbf k)} 
\label{eq:Delta}
\end{equation}
Introducing as in the previous subsection the parity $\eta$ of the state one can write:
\begin{equation}
\beta_{\rm eff}(\mathbf K) = \left[  1 + \frac{2 (1+ \eta) \Lambda^2 G^0_{\rm mol}(\mathbf K)}{g_0} \right]\beta(\mathbf K) 
+\frac{2(1+ \eta) \Lambda^2 g^{\rm mol}}{g_0}   G^0_{\rm mol}(\mathbf K) \chi(\mathbf K)
  \int \frac{d^3K'}{(2\pi)^3} \chi(\mathbf K') G^0_{\rm mol}(\mathbf K') \beta(\mathbf K')  .
\label{eq:betaeff}
\end{equation}
Simple algebra on this expression and comparison of the result 
to Eq.~(\ref{eq:desired_expression}) finally lead to the prescribed substitution Eq.~(\ref{eq:subst}).
An interesting extension of this subsection would be to include a direct interaction between 
the closed-channel molecules and the open-channel atoms, to see if there still exists a closed integral
equation for $D(\kk_1,\kk_2)$.

\section{The 3+1 fermionic problem}
\label{sec:tpofp}

We now consider the four-body problem corresponding to the interaction of three same spin state
fermions of mass $m$ with an extra, distinguishable particle of mass $M$, for example of another
atomic species.
There is no possible $s$-wave interaction among the fermions, and we assume that their interaction
in odd angular momentum waves (in particular the $p$-wave) is negligible.
The fermions thus interact only with the extra particle, either directly in the open channel,
or indirectly through the creation of a closed-channel molecule.
Since the extra particle is alone, there is no need to specify the statistical
nature of that particle, nor of the closed-channel molecule.
An important difference with the previous four-body bosonic case is that the subspace with two closed-channel
molecules will not be populated, and also that there is no direct open channel interaction in the subspace
with two atoms and one closed-channel molecule. All this simplifies the problem.

\subsection{Model Hamiltonian and two-body scattering amplitude}

The fermionic annihilation and creation operators are called $c_\kk$ and $c_\kk^\dagger$, they obey
the usual free space anticommutation relations $\{c_\kk,c_{\kk'}^\dagger\}= (2\pi)^3\delta(\kk-\kk')$.
They are assumed to commute with the extra particle creation operator $a_\kk^\dagger$ and with
a closed-channel molecule creation operator $b_\kk^\dagger$, that also commute among themselves.
The total Hamiltonian $H=H_{\rm at} + H_{\rm mol} + H_{\rm at-mol} + H_{\rm open}$ with
\bea
H_{\rm at} &=& \int \frac{d^3k}{(2\pi)^3} \left[E_\kk c_\kk^\dagger c_\kk + \alpha E_\kk a_\kk^\dagger a_\kk\right] \\
H_{\rm mol} &=& \int \frac{d^3k}{(2\pi)^3} \left(E_{\rm mol} +\frac{\alpha}{1+\alpha} E_\kk \right) b_\kk^\dagger b_\kk \\
H_{\rm at-mol} &=& \Lambda \int \frac{d^3k_1 d^3 k_2}{[(2\pi)^3]^2} 
\chi(\kk_{12})
[b_{\kk_1+\kk_2}^\dagger a_{\kk_1} c_{\kk_2} 
+ a_{\kk_1}^\dagger c_{\kk_2}^\dagger b_{\kk_1+\kk_2}] \\
H_{\rm open} &=& g_0 \int \frac{d^3k_1 d^3 k_2 d^3 k_3 d^3 k_4}{[(2\pi)^3]^4} 
\chi(\kk_{12}) \chi(\kk_{43}) (2\pi)^3 \delta(\kk_1+\kk_2-\kk_3-\kk_4)
a_{\kk_4}^\dagger c_{\kk_3}^\dagger c_{\kk_2} a_{\kk_1}.
\eea
We have introduced the free fermion dispersion relation $E_\kk=\hbar^2 k^2/(2m)$ and the mass
ratio of a fermion to the extra particle:
\be
\alpha \equiv \frac{m}{M}.
\ee
As compared to the bosonic case, we have dropped the factor $1/2$ in front of $g_0$ in $H_{\rm open}$
since the extra particle is distinguishable from the fermions. More important, to maintain the Galilean
invariance, the argument of the cut-off
function $\chi$ is now the relative wave vector of a fermion (of momentum $\hbar\kk_2$) with respect to
the extra particle (of momentum $\hbar \kk_1$):
\be
\kk_{12} \equiv \mu \left(\frac{\kk_2}{m}-\frac{\kk_1}{M}\right)= \frac{\kk_2-\alpha \kk_1}{1+\alpha},
\ee
where $\mu=m M/(m+M)$ is the reduced mass. In this notation for the relative wavevector $\kk_{12}$,
it is important to keep
in mind that the first index refers to the extra particle and the second index to a fermion.

The two-body scattering of the extra particle with a fermion is similar to the two-body bosonic
problem of subsection \ref{subsec:model}. In the bosonic formulas Eqs.~(\ref{eq:phys1},\ref{eq:phys2},\ref{eq:phys3}), 
one simply has to replace $m$ with $2\mu$ and $2\Lambda^2$ with $\Lambda^2$. The two-body $T$ matrix is now given by
\be
\langle \kk_f | T(E+i0^+) |\kk_i\rangle =
-\frac{2\pi\hbar^2}{\mu} \chi(\kk_f)\chi(\kk_i) \, f(E+i0^+)
\ee
where $\kk_i$ and $\kk_f$ are relative wavevectors and the function $f$ now reads
\be
\label{eq:fnow}
\frac{-\mu}{2\pi\hbar^2 f(E+i0^+)} =
\left[g_0 +\frac{\Lambda^2}{E-E_{\rm mol}}\right]^{-1} -\int\frac{d^3k}{(2\pi)^3} \frac{\chi^2(\kk)}{E+i 0^+-\hbar^2 k^2/(2\mu)}.
\ee
The connection of $f$ with the $s$-wave scattering amplitude $f_k$ on shell is still given by
Eq.~(\ref{eq:fk_vs_fE}) with now $k=k_i=k_f=(2\mu E)^{1/2}/\hbar$. Also relation Eq.~(\ref{eq:exprf}) still holds,
with $k=(2\mu E)^{1/2}/\hbar$ for $E\geq 0$ and $k=i(-2\mu E)^{1/2}/\hbar$ for $E<0$.

\subsection{Four-body ansatz and coupled equations}

The ansatz for the four-body state is now simplified to
$|\Psi\rangle=|\psi_{\rm 4\,at}\rangle + |\psi_{\rm 2\,at+1\ mol}\rangle$ with
\bea
|\psi_{4\,\rm at}\rangle &= &
\int \frac{d^3k_1}{(2\pi)^3} \frac{d^3k_2}{(2\pi)^3} \frac{d^3k_3}{(2\pi)^3} \frac{d^3k_4}{(2\pi)^3}
(2\pi)^3 \delta(\kk_1+\kk_2+\kk_3+\kk_4)
A(\kk_1,\kk_2,\kk_3,\kk_4) a_{\kk_1}^\dagger c_{\kk_2}^\dagger c_{\kk_3}^\dagger c_{\kk_4}^\dagger |0\rangle \\
|\psi_{2\,\rm at+1\,mol}\rangle &=&  \int \frac{d^3k_1}{(2\pi)^3} \frac{d^3k_2}{(2\pi)^3} 
B(\kk_1,\kk_2) b_{-(\kk_1+\kk_2)}^\dagger c_{\kk_1}^\dagger c_{\kk_2}^\dagger |0\rangle.
\eea
Taking advantage of the fermionic symmetry, we impose that $A$ is an antisymmetric function of its last three 
vectorial arguments, and that $B$ is an antisymmetric function of its two vectorial arguments.

By projecting Schr\"odinger's equation $0=(H-E)|\Psi\rangle$ in the four-atom sector, where we impose
$E<0$ so as to avoid, as in the bosonic case, vanishing energy denominators,
we obtain
\be
\label{eq:virtue}
0=\int \frac{d^3k_1}{(2\pi)^3} \frac{d^3k_2}{(2\pi)^3} \frac{d^3k_3}{(2\pi)^3} \frac{d^3k_4}{(2\pi)^3}
(2\pi)^3 \delta(\kk_1+\kk_2+\kk_3+\kk_4)
F_4(\kk_1,\kk_2,\kk_3,\kk_4)  a_{\kk_1}^\dagger c_{\kk_2}^\dagger c_{\kk_3}^\dagger c_{\kk_4}^\dagger |0\rangle
\ee
with
\be 
F_4(\kk_1,\kk_2,\kk_3,\kk_4) = \left[-E+\alpha E_{\kk_1} +\sum_{n=2}^{4} E_{\kk_n}\right]A(\kk_1,\kk_2,\kk_3,\kk_4)
+ \chi(\kk_{12}) [3 g_0 \tilde{B}(\kk_3,\kk_4) + \Lambda B(\kk_3,\kk_4)].
\ee
We have defined $\tilde{B}$ by a contraction of $A$ with the cut-off function $\chi$,
\be
\tilde{B}(\kk_3,\kk_4) = \int \frac{d^3k_1 d^3k_2}{[(2\pi)^3]^2} \chi(\kk_{12})  
(2\pi)^3\delta(\kk_1+\kk_2+\kk_3+\kk_4) A(\kk_1,\kk_2,\kk_3,\kk_4).
\ee
Similarly to the bosonic case, the unknown function in the final integral equation will be a linear
combination of $\tilde{B}$ and of $B$:
\be
\Lambda D(\kk_3,\kk_4) \equiv  3 g_0 \tilde{B}(\kk_3,\kk_4) + \Lambda B(\kk_3,\kk_4).
\ee
Since the antisymmetric part of $F_4$ with respect to its last three vectorial arguments is zero by virtue
of Eq.~(\ref{eq:virtue}),
we obtain $A$ as a function of $D$:
\be
\label{eq:Aimp}
A(\kk_1,\kk_2,\kk_3,\kk_4) = \frac{\Lambda/3}{E-(\alpha E_{\kk_1} +\sum_{n=2}^{4} E_{\kk_n})}
\left[
\chi(\kk_{12}) D(\kk_3,\kk_4)
-\chi(\kk_{13}) D(\kk_2,\kk_4)
+\chi(\kk_{14}) D(\kk_2,\kk_3)
\right].
\ee
This we insert in the definition of $\tilde{B}$. After the renumbering $1234\to 3412$ in 
Eq.~(\ref{eq:Aimp}), the extra particle now having the index 3,
we also obtain $\tilde{B}$ as a function of $D$:
\bea
\frac{3}{\Lambda}\tilde{B} (\kk_1,\kk_2) &=& \int \frac{d^3 k_3 d^3 k_4}{[(2\pi)^3]^2} \frac{(2\pi)^3 \delta(\sum_{n=1}^{4} \kk_n)}
{E-(\alpha E_{\kk_3} +\sum_{n\neq 3} E_{\kk_n})}\chi(\kk_{34}) 
 \nonumber \\ 
&&\ \ \ \ \ \ \ \ \ \ \ \ \ \ \ \ \ \ \ \ \ \ \ \ \ \ \ \ \ \ \ \ \ \ \ \ \ \ 
 \times [\chi(\kk_{34}) D(\kk_1,\kk_2) 
 - \chi(\kk_{31}) D(\kk_4,\kk_2) + \chi(\kk_{32}) D(\kk_4,\kk_1)].
\label{eq:impimp1}
\eea

By projecting $0=(H-E)|\Psi\rangle$ in the subspace with two fermions and one closed-channel molecule,
we obtain
\be
0= \int \frac{d^3k_1}{(2\pi)^3} \frac{d^3k_2}{(2\pi)^3} F_2(\kk_1,\kk_2)
b_{-(\kk_1+\kk_2)}^\dagger c_{\kk_1}^\dagger c_{\kk_2}^\dagger |0\rangle
\ee
with
\be
F_2(\kk_1,\kk_2)=\left(-E+E_{\kk_1}+E_{\kk_2}+\frac{\alpha}{1+\alpha} E_{\kk_1+\kk_2} + E_{\rm mol}\right) B(\kk_1,\kk_2)
+3\Lambda \tilde{B}(\kk_1,\kk_2). 
\ee
Since $F_2$ is an antisymmetric function, Schr\"odinger's equation directly imposes $F_2=0$.
As in the bosonic case, we then introduce the relative energy
\be
E_{\rm rel}(\kk_1,\kk_2) \equiv E-\left(E_{\kk_1}+E_{\kk_2}+\frac{\alpha}{1+\alpha} E_{\kk_1+\kk_2}\right),
\ee
and the function
$u(\kk_1,\kk_2) \equiv 1 - (g_0/\Lambda^2)[E_{\rm mol} - E_{\rm rel}(\kk_1,\kk_2)]$.
Then we obtain a second expression of $\tilde{B}$ as a function of $D$:
\be
\label{eq:impimp2}
\frac{3}{\Lambda}\tilde{B}(\kk_1,\kk_2) = g_0^{-1} \left(1-\frac{1}{u(\kk_1,\kk_2)}\right) D(\kk_1,\kk_2).
\ee

\subsection{A single integral equation}

To obtain the desired integral equation on $D$, it remains to eliminate $\tilde{B}$
in between Eq.~(\ref{eq:impimp1}) and Eq.~(\ref{eq:impimp2}).
Diagonal and non-diagonal contributions appear.
In the diagonal contribution, to recover the two-body $T$ matrix, 
we perform the usual change of variables of unit Jacobian, that introduces
the relative wave vector $\kk_{34}$ and the center-of-mass
wave vector $\KK_{34}=\kk_3+\kk_4$  of the extra particle of index 3 and one fermion of index 4:
\bea
\kk_3 &=& \frac{1}{1+\alpha} \KK_{34} - \kk_{34} \\
\kk_4 &=& \frac{\alpha}{1+\alpha} \KK_{34} + \kk_{34}.
\eea
Then, from the identity $(1+\alpha)/m = 1/\mu$, we obtain the simple relation
\be
\frac{(2\pi)^3 \delta(\sum_{n=1}^{4} \kk_n)}
{E-(\alpha E_{\kk_3} +\sum_{n\neq 3} E_{\kk_n})}
=
\frac{(2\pi)^3 \delta (\KK_{34}+\kk_1 +\kk_2)}
{E_{\rm rel}(\kk_1,\kk_2)-\hbar^2 k_{34}^2/(2\mu)}.
\ee
Finally, the integral equation for $D$ in the three fermions plus one extra particle problem is

\hspace{-6.5mm}\fbox{
\parbox[l]{\textwidth}{
\bea
0 &=& \frac{\mu D(\kk_1,\kk_2)}{2\pi\hbar^2 f[E_{\rm rel}(\kk_1,\kk_2)]}
- \int \frac{d^3k_3}{(2\pi)^3} \frac{d^3 k_4}{(2\pi)^3}
\frac{(2\pi)^3 \delta(\sum_{n=1}^{4} \kk_n)}
{E-(\alpha E_{\kk_3} +\sum_{n\neq 3} E_{\kk_n})}
\chi(\kk_{34})
\left[\chi(\kk_{31}) D(\kk_4,\kk_2)\right. \nonumber\\
&&\ \ \ \ \ \ \ \ \ \ \ \ \ \ \ \ \ \ \ \ \ \ \ \ \ \ \ \ \ \ \ \ \ \ \ \ \ \
\ \ \ \ \ \ \ \ \ \ \ \ \ \ \ \ \ \ \ \ \ \ \ \ \ \ \ \ \ \ \ \ \ \ \ \ \ \ \ \ 
\ \ \ \ \ \ \ \ \ \ \ \ \ \ \ \ \ \ \ \ \ \ \ \ 
 \left.+ \chi(\kk_{32}) D(\kk_1,\kk_4)\right],
\label{eq:impfin}
\eea
}}
where the function $f(E)$ is the fermionic one given by Eq.~(\ref{eq:fnow}).
The integral equation (\ref{eq:impfin}) is the main result of this section.
It has to be solved with a fermionic exchange symmetry for $D$, $D(\kk_1,\kk_2)=-D(\kk_2,\kk_1)$.

An instructive limiting case is to take the zero-range limit $b\to 0$ in Eq.~(\ref{eq:impfin}),
assuming that $a_{\rm bg}$ and $R_*$ are $O(b)$ whereas the scattering length $a$ is fixed. 
Then the scattering amplitude of the model tends to the usual zero-range expression
\be
f_k \to -\frac{1}{\frac{1}{a}+ik}. 
\ee
In Eq.~(\ref{eq:impfin}) one replaces the function $\chi$ with unity, and one performs the integral over
the extra particle momentum $\kk_3$. Setting $E=-\hbar^2 q^2/(2\mu)$, $q\geq 0$,
one obtains the integral equation for the zero-range Bethe-Peierls model for three same spin state
fermions and one extra particle:
\bea
&& \left\{\left[q^2+\frac{1+2\alpha}{(1+\alpha)^2}(k_1^2+k_2^2) +\frac{2\alpha}{(1+\alpha)^2}
\kk_1\cdot\kk_2\right]^{1/2}-\frac{1}{a}\right\} D(\kk_1,\kk_2) \nonumber \\
&&\ \ \ \ \ \ \ \ \ \ \ \ \ \ \ \ \ \ \ \ \ \ \ \ \ \ \ \ \ \ \ \ \ \ \ \ \ \ \ \ 
+\int \frac{d^3k_4 }{2\pi^2}
\frac{D(\kk_1,\kk_4)+D(\kk_4,\kk_2)}
{q^2+k_1^2+k_2^2+k_4^2+
\frac{2\alpha}{1+\alpha}
(\kk_1\cdot \kk_2 + \kk_1\cdot \kk_4 + \kk_2\cdot\kk_4)}=0.
\label{eq:minlos}
\eea
This differs from the integral equation derived in \cite{Minlos} by numerical constants
\footnote{In particular, comparison of Eqs.~(10,12) of \cite{Minlos} indicates that a square is missing in the factor $\pi$
of Eq.~(10) in \cite{Minlos}.}.
The zero energy $q\to 0$ and unitary $1/a=0$ limits of Eq.~(\ref{eq:minlos}) were recently used in
\cite{Efimov4} to study the emergence of a pure four-body Efimov effect (without three-body Efimov effect)
in the system of three fermions plus one extra particle, as a function of the mass ratio $\alpha$.

\section{Taking advantage of rotational invariance}
\label{sec:tari}

In practice, integral equations such as Eqs.~(\ref{eq:forme_jolie},\ref{eq:impfin}) are heavy to solve
numerically because the unknown function {\sl a priori} depends on 6 real variables.
A significant reduction of the problem can be achieved by using the rotational symmetry of the Hamiltonian.
This rotational symmetry implies the existence of degenerate eigenenergy
subspaces of well defined total angular momentum $l$. Such a degenerate subspace is associated
to an irreducible representation of the $SO(3)$ rotation group of spin $l$, and it is generated in Dirac's notation
by $2l+1$ functions $|l,m_l\rangle$, with $|l,m_l\rangle$ of angular momentum
$m_l \hbar$ along the arbitrary quantization axis $z$. The general solution $D(\kk_1,\kk_2)$ is a linear
superposition of the particular solutions
\be
\label{eq:defDm}
D_{m_l}(\kk_1,\kk_2) \equiv \langle \kk_1,\kk_2 | l,m_l\rangle.
\ee
In this section we show that a general ansatz for $D_0(\kk_1,\kk_2)$ may be obtained
in terms of $2l+1$ unknown functions of the moduli $k_1$, $k_2$ of $\kk_1$, $\kk_2$,
and of the angle $\theta\in[0,\pi]$ between $\kk_1$ and $\kk_2$. We also consider constraints on these
$2l+1$ functions imposed by the invariance by parity, and by the exchange (bosonic or fermionic) symmetry of
$D(\kk_1,\kk_2)$.

\subsection{Ansatz for an angular momentum $l$}

Consider a general rotation $\mathcal{R}$ of $SO(3)$. In the functional space in which $D(\kk_1,\kk_2)$ lives,
it is represented by the unitary operator $R$, $R^{-1}=R^\dagger$, 
such that 
\be
\label{eq:defR}
R|\kk_1,\kk_2\rangle=|\mathcal{R}\kk_1,\mathcal{R}\kk_2\rangle.
\ee
Within the degenerate eigenenergy subspace of angular momentum $l$, the matrix elements of $R$ form
the unitary matrix $\rho$ of the irreducible representation of $\mathcal{R}$ of spin $l$:
\be
\label{eq:matelemR}
\langle l,m_l|R|l,m_l'\rangle = \rho_{m_l m_l'}.
\ee
This matrix $\rho$ is well known from group theory, where it is expressed by means of the Euler representation
of $\mathcal{R}$ \cite{Wu}:
\be
\label{eq:Euler}
\mathcal{R} = \mathcal{R}_z(\delta_1) \mathcal{R}_y(\delta_2)
\mathcal{R}_z(\delta_3),
\ee
with $\delta_2\in [0,\pi]$, $0\leq \delta_1,\delta_3 < 2\pi$,
and $\mathcal{R}_\nu (\phi)$ in $SO(3)$ is the rotation of angle $\phi$ around axis $\nu$,
$\nu=x,y,z$.

A well-known fundamental relation is then 
\be
\label{eq:invl}
\mathbf{D}\,(\mathcal{R}\kk_1,\mathcal{R}\kk_2) = \rho^*\,
\mathbf{D}\, (\kk_1,\kk_2) = {}^t \rho^{-1} \mathbf{D}\, (\kk_1,\kk_2)
\ee
where the spinor $\mathbf{D}$ is the $2l+1$ vector of components $D_{m_l}$, $\rho^*$ is the complex conjugate
of the matrix $\rho$, ${}^t\rho$ is its transpose, $\rho^{-1}$ its inverse
\footnote{
We recall that Eq.~(\ref{eq:invl}) may be derived by calculating $\langle \kk_1,\kk_2| R^\dagger | l,m_l\rangle$
in two different ways. First, we take $R^\dagger$ to act on the left and we use the definitions
(\ref{eq:defR}) and (\ref{eq:defDm}), so that 
$\langle \kk_1,\kk_2|R^\dagger | l,m_l\rangle = D_{m_l}(\mathcal{R}\kk_1, \mathcal{R}\kk_2)$.
Second, we insert the closure relation $\sum_{m_l'} |l,m_l'\rangle \langle l,m_l'|$ to the left
of $R^\dagger$ in the quantity $\langle \kk_1,\kk_2| R^\dagger | l,m_l\rangle$, and we use Eq.~(\ref{eq:matelemR}).}
.
For {\sl fixed} non-zero wave vectors $\kk_1,\kk_2$, one then choose the matrix $\mathcal{R}$
such that
\bea
\label{eq:Rk1}
\mathcal{R} \kk_1 &=& k_1 \mathbf{e}_x \\
\label{eq:Rk2}
\mathcal{R} \kk_2 &=& k_2 (\cos\theta\, \mathbf{e}_x+\sin\theta\, \mathbf{e}_y),
\eea
where $\theta$ is the non-oriented angle between $\kk_1$ and $\kk_2$, 
\be
\cos\theta = \frac{\kk_1\cdot\kk_2}{k_1 k_2}, \ \  \theta\in [0,\pi],
\ee
and $\mathbf{e}_\nu$ is the unit vector along axis $\nu$, $\nu=x,y,z$. This choice, together with
the fact that $\mathcal{R}$ is a symmetric matrix of determinant unity, implies
\be
\label{eq:Rk12}
\mathcal{R} (\kk_1\wedge\kk_2) = ||\kk_1\wedge\kk_2|| \mathbf{e}_z.
\ee
Remarkably, from Eq.~(\ref{eq:invl}) we see that the spinor $\mathbf{D}$ for arbitrary $\kk_1,\kk_2$ 
is totally determined if one knows the spinor $\mathbf{f}^{(l)}$ which is a function of three real variables only,
\be
\mathbf{f}^{(l)}(k_1,k_2,\theta) \equiv
\mathbf{D}\,[k_1 \mathbf{e}_x,k_2 (\cos\theta\, \mathbf{e}_x+\sin\theta\, \mathbf{e}_y)].
\ee
Eq.~(\ref{eq:invl}) indeed gives
\be
\mathbf{D}\, (\kk_1,\kk_2) = {}^t \rho\, \mathbf{f}^{(l)}(k_1,k_2,\theta).
\ee

To make the ansatz more explicit, we set $D(\kk_1,\kk_2)=D_{m_l'=0}(\kk_1,\kk_2)$ without any loss
of generality, so that $D(\kk_1,\kk_2) = \sum_{m_l=-l}^{l} \rho_{m_l 0} f^{(l)}_{m_l}(k_1,k_2,\theta).$
Using the relation (8.6-1) of \cite{Wu} we find the expression of the matrix element $\rho_{m_l 0}$ in terms of the
spherical harmonics $Y_l^{m_l}(\theta,\phi)$ where $\theta$ is the polar angle and $\phi$ the azimuthal angle.
This leads to
\be
\label{eq:ansatz_general}
D(\kk_1,\kk_2) = \left(\frac{4\pi}{2l+1}\right)^{1/2} 
\sum_{m_l=-l}^{l} \left[Y_l^{m_l} (\delta_2,\delta_1)\right]^* 
f_{m_l}^{(l)}(k_1,k_2,\theta).
\ee

The last step is to calculate the angles $\delta_1$ and $\delta_2$ that enter the Euler representation of $\mathcal{R}$.
From Eq.~(\ref{eq:Euler}) it is apparent that $\delta_1$ and $\delta_2$ are the azimuthal and polar angles of the vector
$\mathcal{R}\mathbf{e}_z$ in the spherical coordinates of axis $z$. The 
Cartesian coordinates
$(X,Y,Z)$ of that vector are readily evaluated
using $\mathbf{e}_\nu \cdot \mathcal{R}\mathbf{e}_z = \mathbf{e}_z\cdot \mathcal{R}^{-1} \mathbf{e}_\nu$,
$\nu=x,y,z$, and the identities Eqs.~(\ref{eq:Rk1},\ref{eq:Rk2},\ref{eq:Rk12}) after their multiplication
by the matrix $\mathcal{R}^{-1}$:
\bea
X &=& \sin\delta_2\cos\delta_1 = \frac{\kk_1\cdot\mathbf{e}_z}{k_1} \\
Y &=& \sin\delta_2\sin\delta_1 = 
\left(\frac{\kk_2\cdot\mathbf{e}_z}{k_2}
-\cos\theta \frac{\kk_1\cdot\mathbf{e}_z}{k_1}\right)/\sin\theta \\
Z &=& \cos\delta_2 = \mathbf{e}_z\cdot\frac{\kk_1\wedge\kk_2}{||\kk_1\wedge\kk_2||}.
\eea
We note that, in the most common case $l=0$, one has $Y_0^0(\theta,\phi)= 1/\sqrt{4\pi}$
so that the general ansatz reduces to the usual form $D(\kk_1,\kk_2)=f^{(0)}_0(k_1,k_2,\theta)$.

\subsection{Inclusion of parity and exchange symmetry}

Let us apply the parity operator to the function $D(\kk_1,\kk_2)$. The change $(\kk_1,\kk_2)\to (-\kk_1,-\kk_2)$
transforms $(X,Y,Z)$ into $(-X,-Y,Z)$ so that the angle $\delta_2$ is unchanged whereas $\delta_1$ is transformed
into $\delta_1+\pi$ modulo $2\pi$. This changes 
$Y_l^{m_l} (\delta_2,\delta_1)$ into $(-1)^{m_l} Y_l^{m_l} (\delta_2,\delta_1)$. This allows to decouple
the even $m_l$ terms of Eq.~(\ref{eq:ansatz_general}), that correspond to an even-parity $\eta=+1$
component of $D$, from
the odd $m_l$ terms of Eq.~(\ref{eq:ansatz_general}), that correspond to an odd-parity $\eta=-1$ component of $D$.
\emph{E.g.}\ if one considers the ansatz of $l=1$ with $\eta=+1$, one has
$Y_1^0(\theta,\phi)= [3/(4\pi)]^{1/2}\cos\theta$ so that 
$D(\kk_1,\kk_2)= \mathbf{e}_z\cdot\frac{\kk_1\wedge\kk_2}{||\kk_1\wedge\kk_2||} f^{(1)}_0(k_1,k_2,\theta).$

Let us now consider the bosonic or fermionic symmetry requirement.
Under the exchange $(\kk_1,\kk_2)\to(\kk_2,\kk_1)$, we find that $(X,Y,Z)$ is transformed into
$(X',Y',Z')$ with
\bea
X' &=& X \cos\theta + Y \sin\theta \\
Y' &=& X \sin\theta -Y \cos\theta \\
Z' &=& -Z.
\eea
Then the polar $\delta_2'$ and azimuthal $\delta_1'$ angles of $(X',Y',Z')$ are given by
$\delta_1'=\theta-\delta_1$ modulo $2\pi$ and $\delta_2'=\pi-\delta_2$.
From the following properties of the spherical harmonics,
\bea
Y_l^m (\pi-\theta,\phi+\pi) &=& (-1)^l Y_l^m(\theta,\phi) \\
Y_l^{-m} (\theta,\phi) &=& (-1)^m Y_l^{m*}(\theta,\phi) \\
Y_l^{m*}(\theta,\phi) &=& Y_l^m(\theta,-\phi),
\eea
we obtain $Y_l^m(\pi-\delta_2,\theta-\delta_1) =
(-1)^l e^{im\theta} Y_l^{-m} (\delta_2,\delta_1).$
The bosonic ($\epsilon=1$) or fermionic ($\epsilon=-1$) symmetry imposes
$D(\kk_2,\kk_1)=\epsilon D(\kk_1,\kk_2)$ so that the unknown functions $f_{m_l}^{(l)}$ in the ansatz
must satisfy
\be
f_{m_l}^{(l)}(k_2,k_1,\theta) = \epsilon (-1)^{l} e^{i {m_l} \theta} f_{-{m_l}}^{(l)}(k_1,k_2,\theta).
\ee
A more convenient form is obtained by the change of unknown functions,
\be
\tilde{f}_{m_l}(k_1,k_2,\theta) \equiv
e^{-i{m_l}\theta/2} f_{m_l}(k_1,k_2,\theta).
\ee
The exchange symmetry then imposes
\be
\tilde{f}_{m_l}(k_2,k_1,\theta) = \epsilon (-1)^{l} \tilde{f}_{-{m_l}}(k_1,k_2,\theta),
\ee
which easily allows to restrict the calculation to the domain $k_2\geq k_1$.

By inserting the general ansatz (\ref{eq:ansatz_general}) into the integral equations
Eq.~(\ref{eq:forme_jolie}) or Eq.~(\ref{eq:impfin}), one may obtain general equations
for the $f_{m_l}^{(l)}$. This was done in Ref.~\cite{Efimov4} in the particular case of the 3+1 fermionic problem
with zero-range interactions, infinite scattering length and at zero energy.
We shall not pursue this general issue here.

\section{Conclusion}
\label{sec:conclusion}

In the frame of a quite realistic two-channel model for the interaction
Hamiltonian close to a magnetic Feshbach resonance, with three physical parameters and a potential range,
we have shown that, for two relevant problems in the current few-body physics with cold
atoms (four bosons, or three same spin state fermions plus an extra particle) it is possible
to derive exactly simple integral equations for an unknown function of six real variables.
Furthermore, using the rotational symmetry of the Hamiltonian and its invariance by parity, 
for an angular momentum $l$, we have shown that the problem may be reduced to integral equations for $l$ 
or $l+1$ unknown functions of 3 real variables only, two wave vector moduli and one angle. 

Within the considered model Hamiltonian,
this makes the {\sl exact} numerical solution of the four-body problem tractable.
We presented elsewhere a first application of this formalism in the limiting case of a zero-range interaction
\cite{Efimov4}. We expect that other applications will appear, in particular in the regimes where the zero-range limit
with no extra three-body parameter is ill-defined, as it is the case for four bosons, or for the $3+1$ fermionic
problem with a fermion to extra particle mass ratio larger than $\simeq 13.607$.



\section*{Acknowledgements}
The group of Y.C.\ is a member of IFRAF. The cold atom group of LPTMC is associated with IFRAF.
We acknowledge useful discussions with F. Werner,
and we thank him for having pointed out the work of Minlos \cite{Minlos}.

\end{document}